\documentclass[aps,twocolumn,showpacs,amsmath,amssymb,superscriptaddress]{revtex4}

\usepackage{graphicx}
\usepackage{dcolumn}
\usepackage{bm}

\begin{document}


\title{Consistent Community Identification in Complex Networks}

\author{Haewoon Kwak}
 \affiliation{Computer Science Department, KAIST, Daejeon, 305-701, Korea}
\author{Young-Ho Eom}%
 \affiliation{Department of Physics, KAIST, Daejeon, 305-701, Korea}
\author{Yoonchan Choi}
 \affiliation{Samsung Advanced Institute of Technology, Gyeonggi, 449-712, Korea}
\author{Hawoong Jeong}%
 \email{hjeong@kaist.edu}
 \affiliation{Department of Physics, KAIST, Daejeon, 305-701, Korea}
 \affiliation{Institute for the BioCentury, KAIST, Dajeon, 305-701, Korea}
\author{Sue Moon}
 \affiliation{Computer Science Department, KAIST, Daejeon, 305-701, Korea}

\date{\today}

\begin{abstract}

We have found that known community identification algorithms
produce inconsistent communities when the node ordering changes at
input. We propose two metrics to quantify the level of consistency
across multiple runs of an algorithm: pairwise membership
probability and consistency. Based on these two metrics, we
address the consistency problem without compromising the
modularity.  Our solution uses pairwise membership probabilities
as link weights and generates consistent communities within six or
fewer cycles. It offers a new tool in the study of community
structures and their evolutions.

\end{abstract}

\pacs{89.75.-k, 89.75.Hc}
\maketitle

Understanding and identifying community structure in a complex
network has been one of the major research topics in sociology,
physics, biology, and computer science~\cite{Fortunato2009}.
Various algorithms for discovering communities and modules in
networks have been proposed: Some are based on betweenness and
similar measures by removing inter-community
links~\cite{Girvan02,Radicchi04}. Others use
cliques~\cite{Palla05}, information theory~\cite{Rosvall07},
random walks on networks~\cite{Rosvall08}, similarity among
partitions~\cite{Gustafsson06}, and the list is not exhausted.

Among these algorithms, greedy modularity maximization is one of
the prevalent approaches for community identification. The
\textit{modularity}, $Q$, is a quality measure of partitioned
communities. It is defined as:
\begin{equation}
Q = \sum_{i} (e_{ii}-a_i^2)
\end{equation}
where $e_{ii}$ is the ratio of the number of links between nodes
belonging to community $i$ over all links and $a_i$ is the ratio
of all links that cross the boundary of community $i$ over all
links.  The value of modularity ranges from -1 to 1.  The value
$Q=0$ implies that the number of links within a community is no
better than random.

Modularity maximization methods (MMMs) are effective in
identifying and uncovering community structure in networked
systems, but they have some limitations. For example, MMMs fail to
identify communities smaller than a certain scale, which is known
as the resolution limit~\cite{Fortunato07}.

In this work we report another limitation of MMMs, namely, the
inconsistency among identified communities in multiple runs of an
algorithm. Using empirical network data, we show that all
algorithms we have reviewed produce inconsistent communities every
time the node names are reordered while the structure of the
network remains unchanged.



\begin{figure*} [hbt!]
  \begin{center}
  \begin{minipage}{54mm}
  \includegraphics[width=52mm]{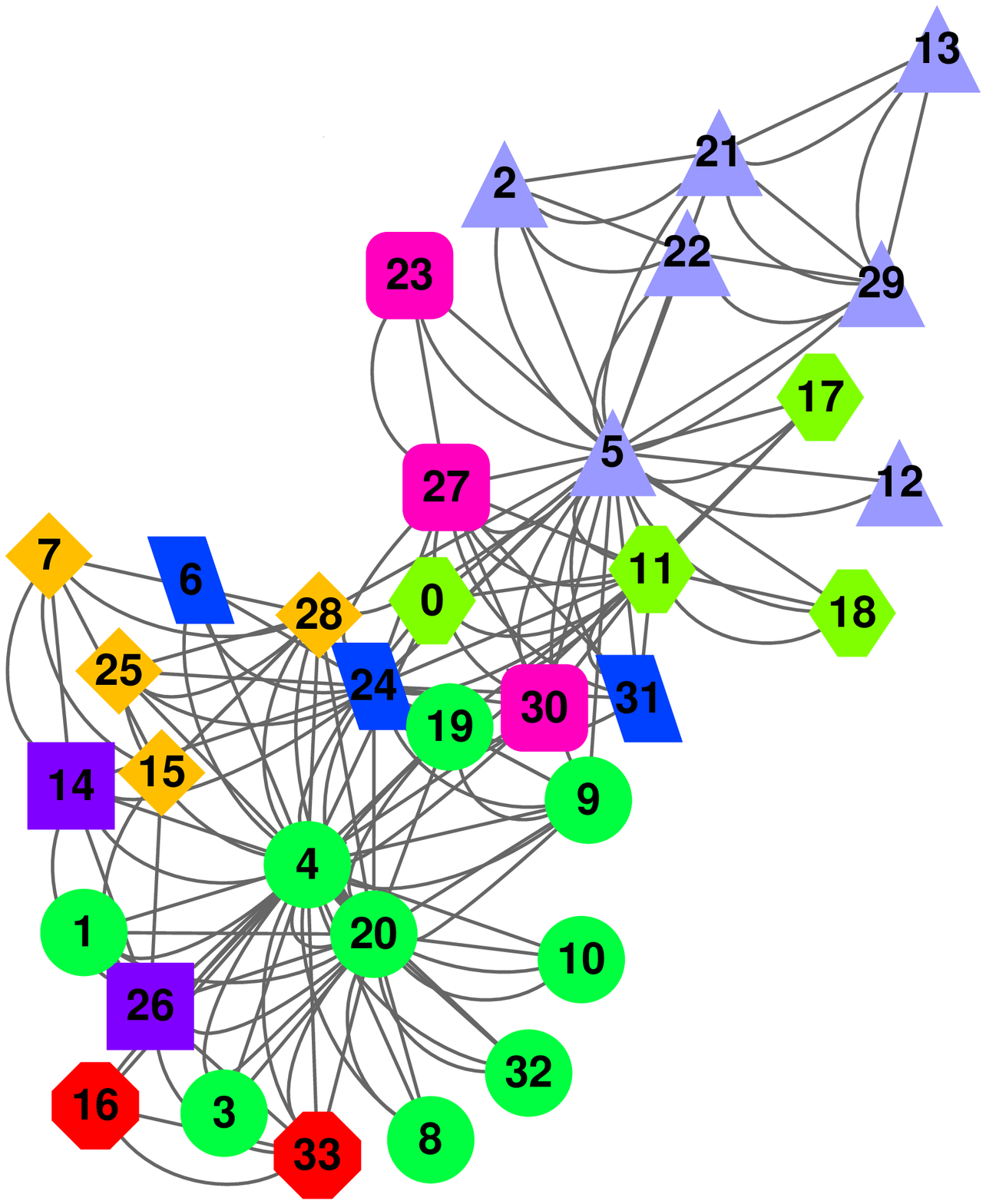}
  \centerline{(a) Q=0.273176}
  \end{minipage}
  \begin{minipage}{54mm}
  \includegraphics[width=52mm]{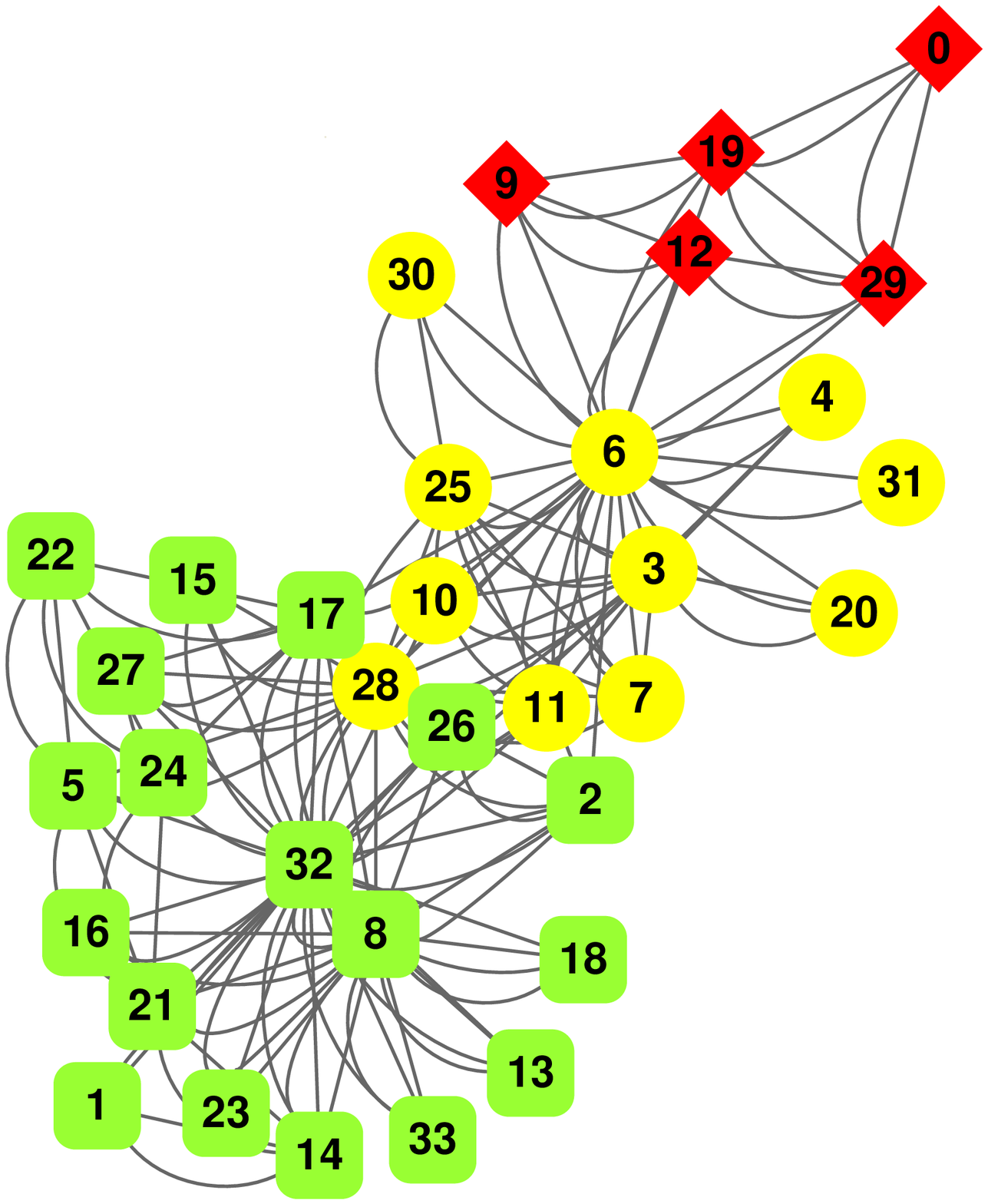}
  \centerline{(b) Q=0.380671}
  \end{minipage}
  \begin{minipage}{54mm}
  \includegraphics[width=52mm]{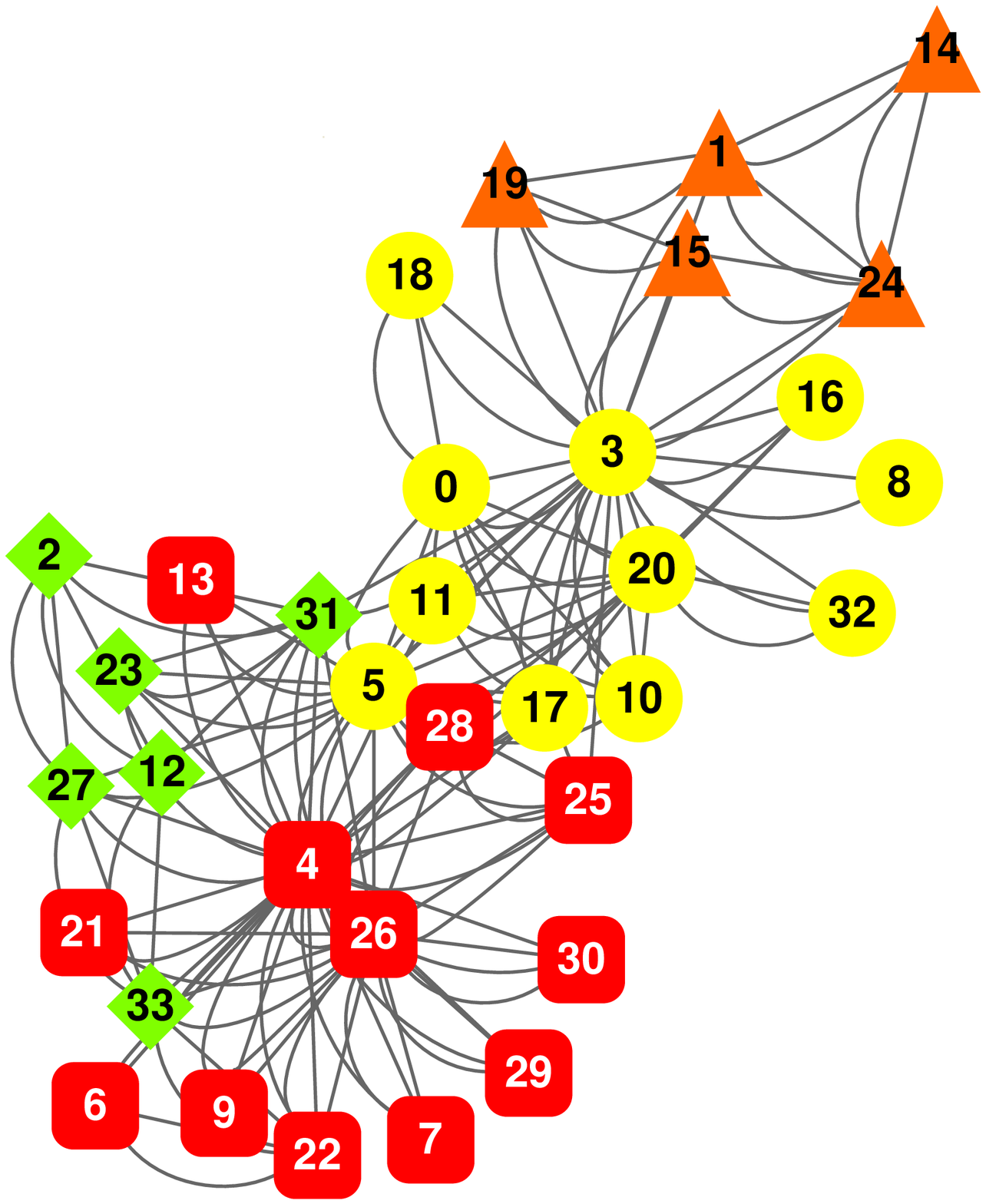}
  \centerline{(c) Q=0.41979}
  \end{minipage}
  \caption{[Color Online] Visualization of inconsistent community
identification in the Karate network~\cite{Zachary77}.  Nodes of a
color belong to the same community, and node ordering is depicted
as the number in the node.}
  \label{fig:communities_in_karate_network}
  \end{center}
\end{figure*}

We consider three community identification algorithms:
Clauset-Newman-Moore (CNM)~\cite{AClauset04},
Wakita~\cite{KWakita07}, and Louvain~\cite{VBlondel08}. They all
take a greedy approach in modularity maximization and are the only
known algorithms to work for large networks. However, they all
produce different values of modularity for the same network. Even
a single algorithm produces different modularities when the input
order of nodes changes. We show an example to illustrate the
inconsistency even in a small well-studied network. The identified
communities in a network by the Louvain algorithm under three
different orderings of nodes are shown in
Fig.~\ref{fig:communities_in_karate_network}. Although the network
has a small number of $34$ nodes, identified communities in
Fig.~\ref{fig:communities_in_karate_network}(a), (b), and (c) are
quite different and have different modularities. This example
demonstrates that even for a small network, the input order plays
a crucial role in determining community structure in complex
networks.

The huge number of ways to partition a graph makes it impossible
to optimize modularity exhaustively. From a macroscopic view this
is fine as long as the modularity varies not too much. However, if
we are interested in network analysis from a nodal perspective,
that is, identifying a community a node belongs to, it does not
make sense for the node to belong to a complete different
community every time the input order is perturbed.  For example,
we have two snapshots of a growing network taken a year apart. How
has the community of a node grown in a year? This question is
about evolutionary clustering, and inconsistent communities are a
problem.  What we address in this work is the inconsistency not
even over the course of evolution, but within a single snapshot.
If the community identification algorithm is so sensitive to the
order of the input and produces completely different communities
from a node's perspective, we cannot answer the question raised in
the example. Thus before we identify the community a node belongs
to, we should ask: how consistent is the community membership
across different input orders?

Over $N$ runs of an algorithm, each with a randomly ordered input
set, we quantify the likelihood of a pair of nodes resulting in
the same community as:
\begin{equation} \label{eq:cmp_matrix}
p_{ij} = \frac{\sum_{n=1}^{N} \delta_{n}(c_i, c_j)}{N}
\end{equation}
where
\begin{displaymath} \label{eq:cm_matrix}
\delta_n(c_i, c_j) = \left\{ \begin{array}{ll}
    $1$, & \textrm{if $c_{i}$ = $c_{j}$ in the $n$th dataset}\\
    $0$, & \textrm{otherwise}
    \end{array} \right.
\end{displaymath}
and $i$ and $j$ are node indices and $c_i$ and $c_j$ represent
communities that $i$ and $j$ belong to, respectively.  We call
this metric \textit{pairwise membership probability}.  The
pairwise membership probability $p_{ij}$ represents the empirical
probability that two nodes belong to the same community across
multiple runs of the same algorithm.  We can compute $p_{ij}$ for
all possible pairs of nodes.  However, for any specific $i$,
$p_{ij}$ is likely to be $0$ for most of $j$ due to the sparsity
of links in the network, and this tendency grows with the network
size.  Therefore, we consider $p_{ij}$ only for those adjacent
nodes; that is, only between neighboring nodes.

The pairwise membership probability of $1$ means that the two
neighboring nodes always belong to the same community and $0$
means that the two never belong to the same community irrespective
of the input order. The larger the number of pairs whose empirical
pairwise membership probability is close to either 0 or 1 is, the
more consistent the identified communities are. While $p_{ij}$
close to $1/2$ means that $i$ and $j$ can be in the same community
more or less randomly.

In order to quantify network-wide community membership
consistency, we define a metric of consistency $\mathcal{C}$ for
the entire network as:
\begin{equation} \label{eq:consistency}
\mathcal{C} = \frac{\displaystyle\sum_{(i, j) \in E}(\displaystyle
p_{ij} - 1/2)^2}{|E|} \times \frac{1}{(0.5)^2}
\end{equation}
and $E$ is the set of links and $|E|$ is the number of links. The
consistency $\mathcal{C}$ weighs the pairwise membership
probabilities away from $1/2$.  The multicative term in
(\ref{eq:consistency}) normalizes $\mathcal{C}$ from 0 to 1.

\begin{figure} [hbt!]
  \begin{center}
  \includegraphics[width=84mm]{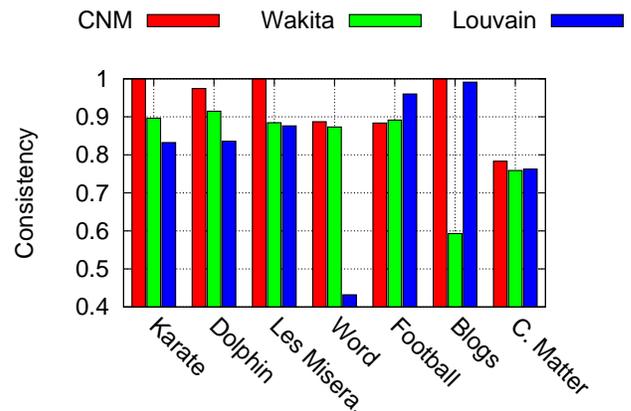}
  \caption{[Color Online] Consistency of community identification}
  \label{fig:comparison_of_consistency}
  \end{center}
\end{figure}

\begin{table*}[t]
\caption{\label{tab:datatable} Summary of the statistics of the
network structure for the three empirical networks. $\textbf{N}$
is the number of nodes, $L$ is the number of links, and $C$ is the
global clustering coefficient. }
\begin{center}
\begin{tabular}{cccccccc}
\hline \hline & Karate & Dolphin & Les  & Word  & Football & Political & Condensed \\
 & & & miserables & adjacencies & & blog &  \\
 \hline
$\textbf{N}$ & 34 & 62 & 77 & 112 & 115 & 1222 & 36458 \\
$L$ & 78 & 159 & 254 & 425 & 613 & 16714 & 171736 \\
$\langle k \rangle$ & 4.6& 5.1 & 6.6 & 7.6 & 10.7 & 27.4 & 9.4 \\
$C$ & 0.57& 0.26 & 0.57 & 0.17 & 0.40 & 0.32 & 0.66\\
\hline \hline \label{tab:dataset}
\end{tabular}
\end{center}
\end{table*}

We have analyzed consistency in community memberships of seven
empirical systems from various fields such as the karate
club~\cite{Zachary77}, dolphin social network~\cite{Lusseau2003},
the co-appearance network of characters in the novel Les
Miserables~\cite{Knuth1993}, the adjacency network of common
adjectives and nouns in the novel David
Copperfield~\cite{Newman2006}, the regular season network of
American football games between Division IA colleges during the
Fall 2000~\cite{Girvan02}, a directed network of hyperlinks
between weblogs on US politics~\cite{Adamic2005} and the network
of coauthorships between scientists posting preprints on the
Condensed Matter E-Print Archive~\cite{Newman2001}.
Table~\ref{tab:dataset} shows basic statistics of the seven
networks.

In case of communities detected by the CNM algorithm in the Karate
club, $12.8$\% of the pairwise membership probabilities are $0$
and the rest of the pairs have $1$, which means that nodes of a
community always belong to the same community over $N$ runs:
$\mathcal{C} = 1$.  In Fig.~\ref{fig:comparison_of_consistency} we
show the consistency from the three algorithms. There is no one
algorithm that outperforms the other two in all networks and no
consistent correlation between the consistency and the topological
characteristics of the network, such as network size, average
degree and average clustering coefficient. However, a closer look
at pairwise membership probabilities reveals that in all networks
far more than $50$\% of pairs have pairwise membership
probabilities either smaller than $0.2$ or greater than
$0.8$~\cite{Kwak2009}. It means that most pairs of nodes are never
in the same community or always in the same community,
respectively. Based on this observation, we devise a consistency
reinforcing mechanism as follows.  After each \textit{cycle} of
$N$ runs, we calculate the pairwise membership probabilities and
then assign them as link weights.  From the second cycle on, we
use this weighted network as an input and continue the cycle until
$\mathcal{C}$ reaches $0.999$ or higher. In a weighted network, an
edge of a higher weight is placed within a community, while an
edge of a lower weight bridges communities. Since we assign the
pairwise membership probability as the weight of the corresponding
link, an edge of high pairwise membership probability in the prior
cycle is more likely to be placed within a community in the next
cycle. Therefore, links with higher weights are reinforced through
multiple cycles and eventually consistent communities emerge.

Our approach has the effect of removing those links with pairwise
membership probabilities of $0$ in the next cycle and spreading
unit link weight between $0$ and $1$, thus reducing ties
significantly in calculating $\Delta Q$.  When there are ties, can
we give preference to nodes based on other metrics, such as
degrees or betweenness centrality\cite{Girvan02}?  To assess the
benefit of other metrics, if any, we order nodes by the degree,
clustering coefficient, degree correlation, and betweenness
centrality and compute modularity.
Even if we employ all the metrics in tie breaking, we cannot
eliminate ties completely~\cite{Kwak2009}. In other words, no
single topological characteristic consistently stands out to work
better than others in all networks.  We have looked at edge
betweenness as well, and found no correlation between edge
betweenness and pairwise membership probability.

Our approach of reinforcing consistency in multiple cycles is
applicable to any of the three algorithms.  We include only the
results from the Louvain algorithm in this paper, for it is the
fastest and only one that scales up to billions of links. We
report that the other two algorithm have similar results.

\begin{figure}
  \begin{center}
  \includegraphics[width=84mm]{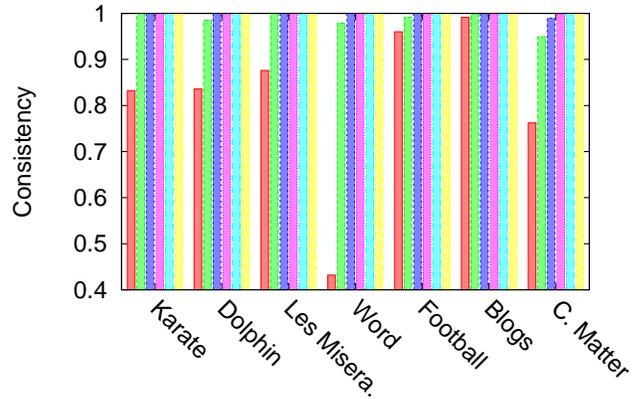}
  \caption{Convergence of consistency }
  \label{fig:convergence_of_consistency}
  \end{center}
\end{figure}

\begin{figure}[tb!]
\begin{center}

\begin{minipage}{66mm}
\includegraphics[width=66mm]{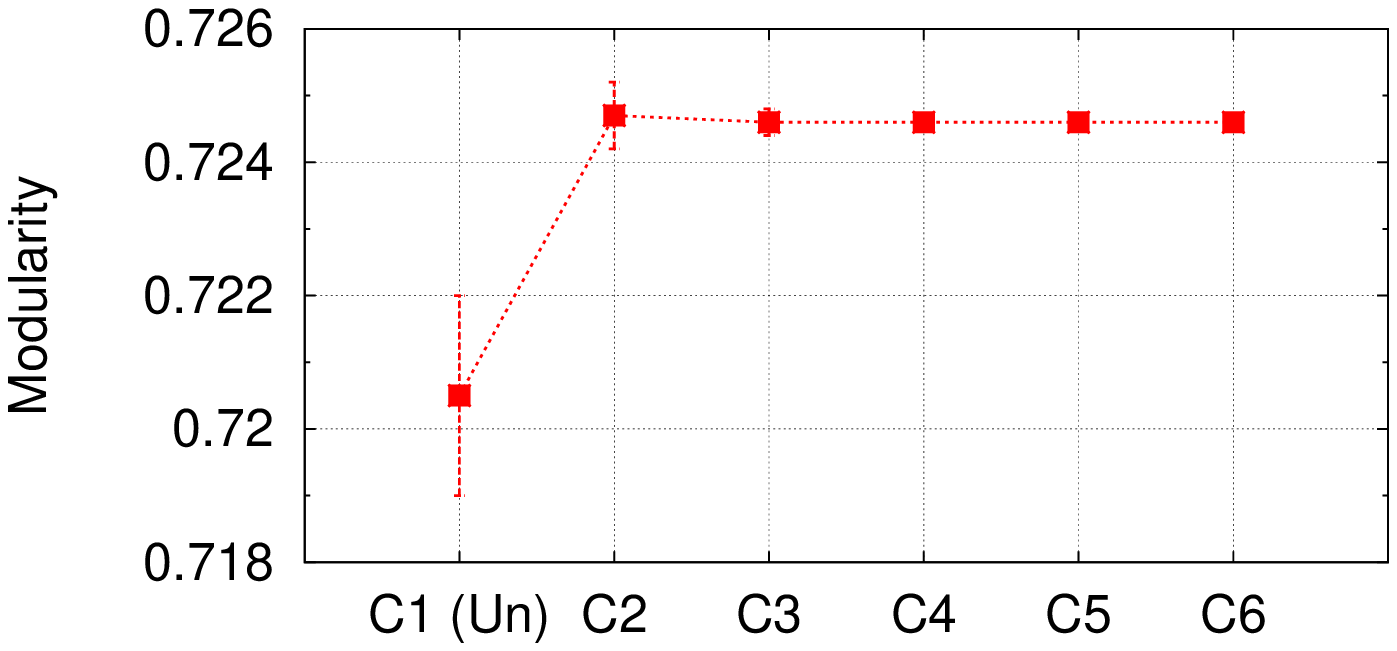}\\
\centerline{(a) Condensed matter}
\end{minipage}
\begin{minipage}{66mm}
\includegraphics[width=66mm]{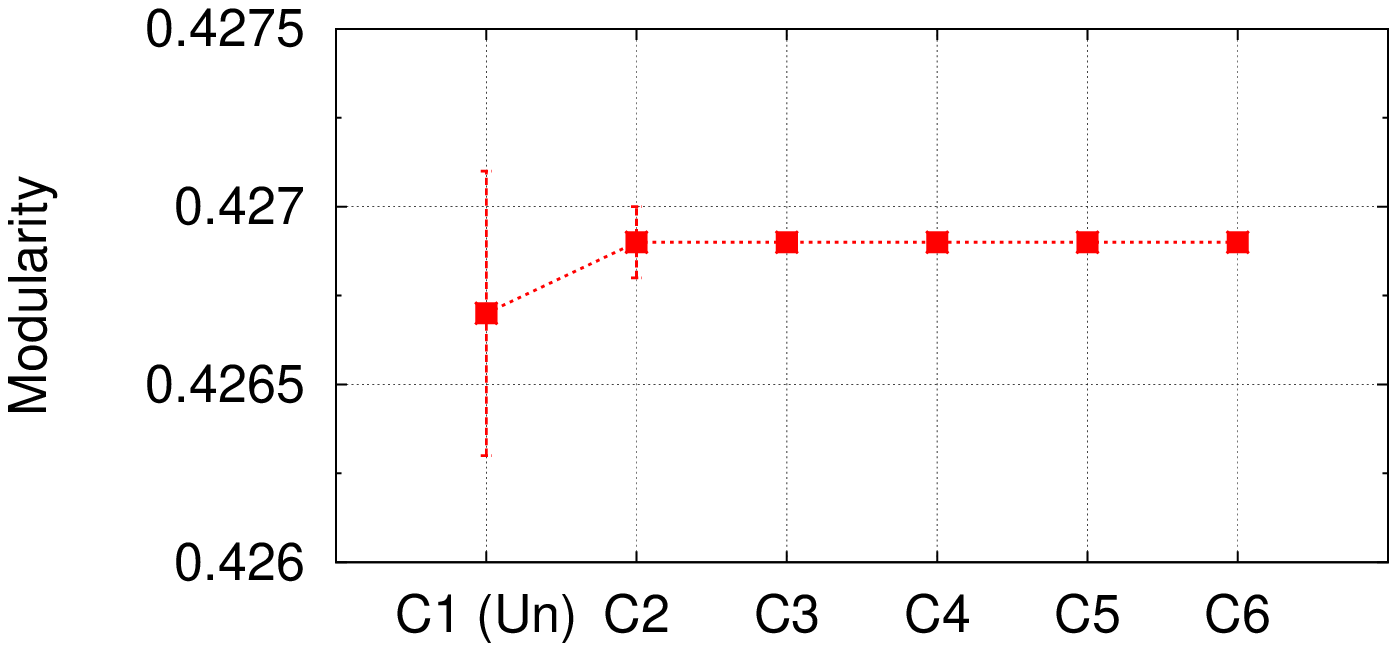}\\
\centerline{(b) Political blogs}
\end{minipage}
\caption{[Color Online] Convergence of modularity ('Un' indicates
modularity of unweighted network)}
\label{fig:convergence_of_modularity}
\end{center}
\end{figure}

The convergence of consistency after $5$ cycles is shown in
Fig.~\ref{fig:convergence_of_consistency}.  All networks
consistency reaches $1$ in $5$ cycles. In
Fig.~\ref{fig:convergence_of_modularity} we show how the
modularity converges over $5$ cycles. The modularity converges
almost to a single point after $2$ cycles.  Furthermore, the
modularity after convergence is higher.
Figure~\ref{fig:convergence_of_modularity} demonstrates that our
approach has no negative impact on modularity, and even improves
it in certain networks.

So far we have shown that our solution of using pairwise
membership probabilities as link weights has improved consistency
greatly. Now we check if communities from different trials come
out identically. We turn our focus to individual communities in
two independent trials. A cycle is $N$ runs for a given network. A
trial is $M$ cycles of a given ordering of the network. We use
$M=6$ and $N=100$. In order to check if the communities are
identical across trials, we calculate the maximum Jaccard
coefficient (the ratio of intersection to union of two
communities) of a community against all communities of another
trial. The Jaccard coefficient of 1 means that the same
communities are produced in both trials. We compare the Jaccard
coefficients for all pairs of trials and most Jaccard coefficients
are found to be greater than 0.95.

In summary, we have investigated the inconsistency among
communities by existing community identification algorithms:
CNM~\cite{AClauset04}, Wakita~\cite{KWakita07}, and
Louvain~\cite{VBlondel08}.  Using empirical network data, we have
shown that all three algorithms produce inconsistent communities
every time the node ordering changes even if the size of networks
are small. Similar results based on very large online social
networks are also reported~\cite{Kwak2009}. To quantify
consistency of identified communities, we introduced pairwise
membership probability and consistency. The former quantifies the
likelihood of two nodes resulting in the same community, and the
latter represent the global level of consistency of a network,
derived from pairwise membership probabilities.  We analyze seven
empirical networks in terms of the above two metrics and show that
no one algorithm outperforms the other two in all networks.
However, most pairwise membership probabilities are close to
either $0$ or $1$ (that is, never in the same community or always
in the same community, respectively). Based on this observation,
we have proposed a solution that improves the consistency without
compromising the modularity.  The key idea is to set the pairwise
membership probability as the link weight and find communities in
the weighted network iteratively. We have demonstrated the
convergence of consistency within $6$ or fewer cycles. Resulting
communities exhibit consistent grouping through multiple trials.\\

\begin{acknowledgments}
This work was supported by NAP of Korea Research Council of
Fundamental Science and Technology and by Basic Science Research
Program through the NRF of Korea funded by the Ministry of
Education, Science and Technology (2009-0087691).
\end{acknowledgments}


\end{document}